\documentclass[sigconf]{acmart}
\pdfoutput=1

\acmConference[TD’15]{Technical Data Conference}{November 12--16}{Dallas, TX, USA}

\title{Persistent Memory I/O Primitives}

\author{Alexander van Renen}
\orcid{0000-0002-6365-4592}
\affiliation{\institution{Technische Universit\"at M\"unchen}}
\email{renen@in.tum.de}
\author{Lukas Vogel}
\affiliation{\institution{Technische Universit\"at M\"unchen}}
\email{vogell@in.tum.de}
\author{Viktor Leis}
\orcid{0000-0001-5676-8017}
\affiliation{\institution{Friedrich-Schiller-Universit\"at Jena}}
\email{viktor.leis@uni-jena.de}
\author{Thomas Neumann}
\affiliation{\institution{Technische Universit\"at M\"unchen}}
\email{neumann@in.tum.de}
\author{Alfons Kemper}
\affiliation{\institution{Technische Universit\"at M\"unchen}}
\email{kemper@in.tum.de}

\begin{abstract}
		I/O latency and throughput is one of the major performance bottlenecks for disk-based database systems.
		Upcoming persistent memory (PMem) technologies, like Intel's Optane DC Persistent Memory Modules, promise to bridge the gap between NAND-based flash (SSD) and DRAM, and thus eliminate the I/O bottleneck.
		In this paper, we provide one of the first performance evaluations of PMem in terms of bandwidth and latency.
		Based on the results, we develop guidelines for efficient PMem usage and two essential I/O primitives tuned for PMem: log writing and block flushing.
\end{abstract}

\usepackage{listings}
\usepackage[utf8]{inputenc}
\usepackage{microtype}

\usepackage{tikz,fontawesome}
\usepackage{pgfplots}
\usepackage{tikz}
\usepackage{cleveref}
\usepackage[binary-units]{siunitx}
\usepackage{todonotes}
\usepackage{epigraph}
\usepackage{makecell}
\usepackage[position=top]{subcaption}
\usepackage{colortbl}
\usepackage{arydshln}
\usepackage{lstautogobble}  
\usepackage{graphicx}
\usepackage{enumitem}
\setitemize{noitemsep,topsep=0pt,parsep=0pt,partopsep=0pt,leftmargin=20pt}
\usepackage{balance}

\usetikzlibrary{intersections}
\usetikzlibrary{arrows.meta}
\usetikzlibrary{decorations.markings}
\usetikzlibrary{patterns}
\pgfplotsset{compat=1.14}
\usepgfplotslibrary{fillbetween}

\makeatletter
\let\myPercent\@percentchar
\makeatother

\makeatletter
\tikzset{
	nomorepostactions/.code={\let\tikz@postactions=\pgfutil@empty},
	mymark/.style n args={3}{decoration={markings,
			mark= between positions 0 and 1 step (1/#3)*\pgfdecoratedpathlength with{%
				\tikzset{#2,every mark}\tikz@options
				\pgfuseplotmark{#1}%
			},
		},
		postaction={decorate},
		/pgfplots/legend image post style={
			mark=#1,mark options={#2},every path/.append style={nomorepostactions}
		},
	},
}
\makeatother

\DeclareSIUnit{\million}{\text{M}}
\DeclareSIUnit{\thousand}{\text{K}}

\definecolor{color1}{rgb}{0.215686,0.494118,0.721569} 
\definecolor{color2}{rgb}{0.894118,0.101961,0.109804} 
\definecolor{color3}{rgb}{0.201961,0.586275,0.190196} 
\definecolor{color4}{rgb}{0.1,0.1,0.1} 
\definecolor{color5}{HTML}{371BE4} 
\definecolor{color6}{rgb}{0.527,0.2009,0.0} 

\definecolor{coldPage}{HTML}{56A5EC}
\definecolor{warmPage}{HTML}{ED9C55}
\definecolor{hotPage}{HTML}{C11B17}

\definecolor{colorMix1}{rgb}{0.667974, 0.23268, 0.3137256667}
\definecolor{colorMix2}{rgb}{0.44183, 0.363399, 0.5176473333}
\definecolor{color1Bright}{rgb}{0.515686,0.794118,0.971569}
\definecolor{color1Dark}{rgb}{0.115686,0.394118,0.621569}
\definecolor{color3Dark}{rgb}{0.101961,0.406275,0.100196}

\tikzstyle{alexArrow} = [-{Straight Barb[length=2mm,width=2mm]}]
\tikzstyle{-alexArrow} = [-{Straight Barb[length=2mm,width=2mm]}]
\tikzstyle{alexArrow-} = [{Straight Barb[length=2mm,width=2mm]}-]

\newcommand{\nvmLine}{\tikz[baseline=-0.0ex]{\draw[color4, thick] (0,0.1) -- plot[mark=x] coordinates {(0.2,0.1)} -- (0.4,0.1);}}

\newcommand{\tierNoOptLine}{\tikz[baseline=-0.0ex]{\draw[color2, thick] (0,0.1) -- plot[mark=o] coordinates {(0.2,0.1)} -- (0.4,0.1);}}

\tikzstyle{alexNumber} = [{text centered, font={\sffamily}, draw, circle, fill=color1Bright, inner sep=1}]

\newcommand{\alexNumberNoneRef}[1]{(#1)}

\newcommand{\beforeCaptionSpacing}{\vspace{-0.2cm}}
\newcommand{\afterCaptionSpacing}{\vspace{-0.2cm}}

\definecolor{bluekeywords}{rgb}{0.13, 0.13, 1}
\definecolor{greencomments}{rgb}{0.1, 0.45, 0.1}
\definecolor{redstrings}{rgb}{0.9, 0, 0}
\definecolor{graynumbers}{rgb}{0.5, 0.5, 0.5}

\begin{document}

\clubpenalty = 10000
\widowpenalty = 10000

\copyrightyear{} 
\acmYear{} 
\setcopyright{acmlicensed}
\acmConference[]{}{}{}
\acmBooktitle{}
\acmPrice{}
\acmDOI{}
\acmISBN{}

\maketitle
\renewcommand{\shortauthors}{Alexander van Renen et al.}

\section{Introduction}

Today, data management systems mainly rely on solid state drives (NAND flash) or magnetic disks to store data.
These storage technologies offer persistence and large capacities at low cost.
However, due to the high access latencies, most systems also use volatile main memory in the form of DRAM as a cache.
This yields the traditional two-layered architecture, as DRAM cannot solely be used due to its volatility, high cost, and limited capacity.

Novel storage technologies, such as Phase Change Memory, are about to shrink this fundamental gap between memory and storage.
Specifically, Intel's upcoming \textit{Optane DC Persistent Memory Modules} (Optane DC PMM) offer an amalgamation of the best properties of memory and storage---though as we show in this paper, with some trade-offs.
This Persistent Memory (PMem) is durable, like storage, and directly addressable by the CPU, like memory.
We also expect the price, capacity, and latency to lie between DRAM and flash.

PMem promises to greatly improve the latency of storage technologies, which in turn would greatly increase the performance of data management systems.
However, because PMem is fundamentally different from existing, well-known technologies, it also has different performance characteristics to DRAM and flash.
In this work, we show how to efficiently implement atomic log writing and page flushing---two critical I/O primitives for database systems.
While we perform our evaluation in a database context, these two I/O primitives are transferable to other systems, as evidenced by the fact that they are also implemented by the Persistent Memory Development Kit (PMDK)~\cite{PMDK}.
The results reported are based on a prototype of Intel's Optane DC PMM rather than software or hardware-based emulation.
Our contributions can be summarized as follows:

\begin{itemize}
\item
We provide one of the first analyses of PMem on a prototype of Intel's Optane DC PMM.
We highlight the impact of the physical properties of PMem on software and derive guidelines for efficient usage of PMem.

\item
We introduce an algorithm for persisting small data chunks (transactional log entries) that reduces the latency by $2\times$ compared to state-of-the-art algorithms.

\item
We investigate different algorithms for persisting large data chunks (database pages) in a failure atomic fashion to PMem.
By combining a copy-on-write method with temporary delta files, we achieve significant speedups.
\end{itemize}

\section{PMem Characteristics}
\label{sec:micro_benchmarks}


\begin{figure*}
	\centering
	\ref{ledgend:bandwidth_st}
	\\
	\vspace{2mm}
	\subcaptionbox{PMem write bandwidth} {
		\hspace{-8mm}
		\begin{tikzpicture}
   \begin{axis}[
      width=\textwidth/3.5,
      height=3.5cm,
      xlabel={\# adjacent cache lines\vphantom{[]}},
      ylabel={\si{\giga\byte/\second}},
      axis lines=left,
      ymin=0, ymax=7.5,
      xmin=1, xmax=12.5,
      xtick={2,4,6,8,10,12},
      ytick={0,2,4,6},
      legend columns=4,
      legend cell align={left},
      legend to name={ledgend:bandwidth_st},
      legend style={at={(0.5,0.8)},anchor=west},
      ]
      \legend{streaming store, store+clwb, store, load}

      \addplot[samples=1, thick, domain=0:6, dotted, name path=three, forget plot] coordinates {(4,0)(4,6.5)};
      \addplot[samples=1, thick, domain=0:6, dotted, name path=three, forget plot] coordinates {(8,0)(8,5.7)};
      \addplot[samples=1, thick, domain=0:6, dotted, name path=three, forget plot] coordinates {(12,0)(12,5.8)};
      
      \node[right, color2] at (4.2,6.8) {\small{6.8 \si{\giga\byte/\second}}};

      \addplot[thick,color=color2,mark=o] file {figures/bandwidth/scale_cls/data/write_nvm_stream.data};
      \addplot[thick,color=color4,mark=x] file {figures/bandwidth/scale_cls/data/write_nvm_clwb.data};
      \addplot[thick,color=color1,mark=triangle] file {figures/bandwidth/scale_cls/data/write_nvm_scalar.data};
      \addplot[thick,color=color3,mark=square] plot coordinates {(100, 0)};

   \end{axis}
\end{tikzpicture}
	}
	\hspace{2mm}
	\subcaptionbox{DRAM write bandwidth} {
		\hspace{-8mm}
		\begin{tikzpicture}
   \begin{axis}[
      width=\textwidth/3.5,
      height=3.5cm,
      xlabel={\# adjacent cache lines\vphantom{[]}},
      axis lines=left,
      ymin=0, ymax=90,
      xmin=1, xmax=12.5,
      ytick={0,20,40,60,80},
      xtick={2,4,6,8,10,12},
      ]

      \addplot[thick,color=color2,mark=o] file {figures/bandwidth/scale_cls/data/write_ram_stream.data};
      \addplot[thick,color=color4,mark=x] file {figures/bandwidth/scale_cls/data/write_ram_clwb.data};
      \addplot[thick,color=color1,mark=triangle] file {figures/bandwidth/scale_cls/data/write_ram_scalar.data};
      
      \node[right, color2] at (8,69) {\small{80 \si{\giga\byte/\second}}};

   \end{axis}
\end{tikzpicture}
	}
	\hspace{2mm}
	\subcaptionbox{PMem read bandwidth} {
		\hspace{-8mm}
		\begin{tikzpicture}
   \begin{axis}[
      width=\textwidth/3.5,
      height=3.5cm,
      xlabel={\# adjacent cache lines\vphantom{[]}},
      axis lines=left,
      ymin=0, ymax=46,
      xmin=1, xmax=12.5,
      xtick={2,4,6,8,10,12},
      ytick={0,10,20,30,40},
      ]

      \addplot[samples=1, thick, domain=0:6, dotted, name path=three, forget plot] coordinates {(4,0)(4,37.5)};
      \addplot[samples=1, thick, domain=0:6, dotted, name path=three, forget plot] coordinates {(8,0)(8,37.5)};
      \addplot[samples=1, thick, domain=0:6, dotted, name path=three, forget plot] coordinates {(12,0)(12,35.0)};

      \node[left, color3, fill=white] at (11.6,14.2) {\footnotesize w/ prefetcher};

      \node[right, color3] at (4.2,37.0) {\footnotesize{37 \si{\giga\byte/\second}}};

      \addplot[thick,color=color3,mark=square] file {figures/bandwidth/scale_cls/data/read_nvm.data};
      \addplot[thick,color=color3,mark=square] file {figures/bandwidth/scale_cls/data/read_nvm_with_prefetcher.data};

   \end{axis}
\end{tikzpicture}
	}
	\hspace{2mm}
	\subcaptionbox{DRAM read bandwidth} {
		\hspace{-8mm}
		\begin{tikzpicture}
   \begin{axis}[
      width=\textwidth/3.5,
      height=3.5cm,
      xlabel={\# adjacent cache lines\vphantom{[]}},
      axis lines=left,
      ymin=0, ymax=120,
      xmin=1, xmax=12.5,
      ytick={0,20,40,60,80,100},
      xtick={2,4,6,8,10,12},
      ]

      \node[left, color3] at (11.5,68.0) {\footnotesize w/ prefetcher};

      \addplot[thick,color=color3,mark=square] file {figures/bandwidth/scale_cls/data/read_ram.data};
      \addplot[thick,color=color3,mark=square] file {figures/bandwidth/scale_cls/data/read_ram_with_prefetcher.data};

      \node[right, color3] at (7,112) {\footnotesize{101 \si{\giga\byte/\second}}};

   \end{axis}
\end{tikzpicture}
	}
\beforeCaptionSpacing
\caption{\textbf{PMem Bandwidth: Varying Access Granularity} -- PMem bandwidth (\textbf{a}, \textbf{c}) with \SI{24}{\textrm{threads}} compared to DRAM bandwidth (\textbf{b}, \textbf{d}) with a varying number of adjacently accessed cache lines. We use a random access pattern that allows for out-of-order execution.}
\afterCaptionSpacing
\label{fig:bandwidth_scale_cls}
\end{figure*}
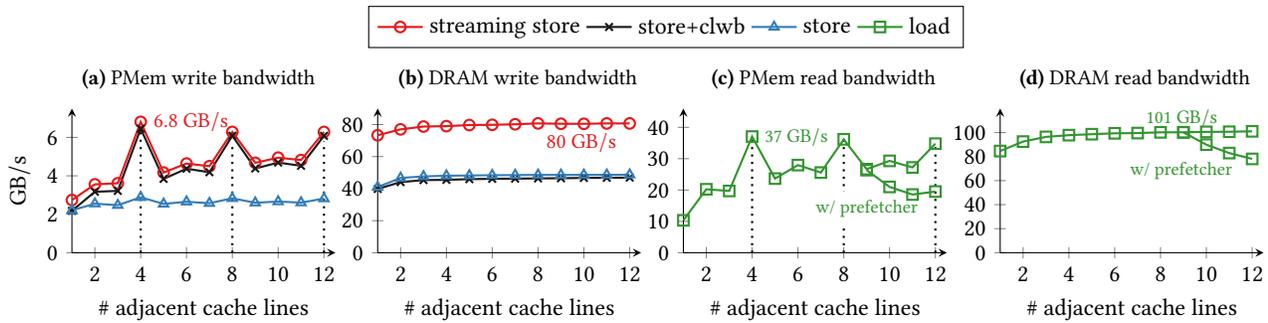


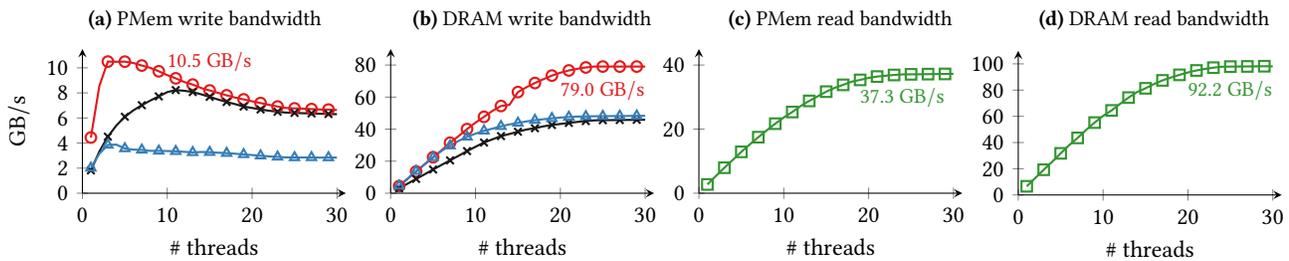
\begin{figure*}
	\centering
	\vspace{2mm}
	\subcaptionbox{PMem write bandwidth} {
		\hspace{-8mm}
		\begin{tikzpicture}
   \begin{axis}[
	  width=\textwidth/3.5,
	  height=3.5cm,
      xlabel={\# threads\vphantom{[]}},
      ylabel={\si{\giga\byte/\second}},
      axis lines=left,
      ymin=0, ymax=11.5,
      xmin=0, xmax=31.05,
      ytick={0,2,4,6,8,10},
      mark repeat = 2,
      mark phase = 1,
      legend columns=4,
      legend cell align={left},
      legend to name={ledgend:bandwidth_mt},
      legend style={at={(0.5,0.8)},anchor=west},
      ]
      \legend{streaming store, store+clwb, store, load}

      \addplot[thick,color=color2,mark=o] file {figures/bandwidth/scale_threads/data/write_nvm_stream.data};
      \addplot[thick,color=color4,mark=x] file {figures/bandwidth/scale_threads/data/write_nvm_clwb.data};
      \addplot[thick,color=color1,mark=triangle] file {figures/bandwidth/scale_threads/data/write_nvm_scalar.data};
      \addplot[thick,color=color3,mark=square] plot coordinates {(100, 0)};

      \node[right, color2] at (9,10.5) {\small{10.5 \si{\giga\byte/\second}}};

   \end{axis}
\end{tikzpicture}
	}
	\hspace{2mm}
	\subcaptionbox{DRAM write bandwidth} {
		\hspace{-8mm}
		\begin{tikzpicture}
   \begin{axis}[
      width=\textwidth/3.5,
      height=3.5cm,
      xlabel={\# threads\vphantom{[]}},
      axis lines=left,
      ymin=0, ymax=90,
      xmin=0, xmax=31.05,
      mark repeat = 2,
      mark phase = 1,
      ytick={0,20,40,60,80},
      ]

      \addplot[thick,color=color2,mark=o] file {figures/bandwidth/scale_threads/data/write_ram_stream.data};
      \addplot[thick,color=color4,mark=x] file {figures/bandwidth/scale_threads/data/write_ram_clwb.data};
      \addplot[thick,color=color1,mark=triangle] file {figures/bandwidth/scale_threads/data/write_ram_scalar.data};

      \node[right, color2] at (19, 64) {\small{79.0 \si{\giga\byte/\second}}};

   \end{axis}
\end{tikzpicture}
	}
	\hspace{2mm}
	\subcaptionbox{PMem read bandwidth} {
		\hspace{-8mm}
		\begin{tikzpicture}
   \begin{axis}[
      width=\textwidth/3.5,
      height=3.5cm,
      xlabel={\# threads\vphantom{[]}},
      axis lines=left,
      ymin=0, ymax=45,
      xmin=0, xmax=31.05,
      mark repeat = 2,
      mark phase = 1,
      ]

      \addplot[thick,color=color3,mark=square] file {figures/bandwidth/scale_threads/data/read_nvm.data};

      \node[right, color3] at (18,30) {\small{37.3 \si{\giga\byte/\second}}};

   \end{axis}
\end{tikzpicture}
	}
	\hspace{2mm}
	\subcaptionbox{DRAM read bandwidth} {
		\hspace{-8mm}
		\begin{tikzpicture}
   \begin{axis}[
      width=\textwidth/3.5,
      height=3.5cm,
      xlabel={\# threads\vphantom{[]}},
      axis lines=left,
      ymin=0, ymax=110,
      xmin=0, xmax=31.05,
      ytick={0,20,40,60,80,100},
      mark repeat = 2,
      mark phase = 1,
      ]

      \addplot[thick,color=color3,mark=square] file {figures/bandwidth/scale_threads/data/read_ram.data};

      \node[right, color3] at (19,80) {\small{92.2 \si{\giga\byte/\second}}};

   \end{axis}
\end{tikzpicture}
	}
\beforeCaptionSpacing
\caption{\textbf{PMem Bandwidth: Varying Thread Count} -- PMem bandwidth (\textbf{a}, \textbf{c}) compared to DRAM bandwidth (\textbf{b}, \textbf{d}) for \num{4} adjacent cache lines with an increasing number of threads. We use a random access pattern that allows for out-of-order execution.}
\afterCaptionSpacing
\label{fig:bandwidth_scale_threads}
\end{figure*}

In this section, we first describe how we configured our system before presenting latency and bandwidth results.

\subsection{Setup and Configuration}

There are two ways of using PMem: memory mode and app direct mode.
In \textbf{memory mode}, PMem replaces DRAM as the (volatile) main memory, and DRAM serves as an additional hardware managed caching layer (``L4 cache'').
The advantage of this mode is that it works transparently for legacy software and thus offers a simple way of extending the main memory capacity at low cost.
However, this does not utilize persistence, and performance may degrade due to the lower bandwidth and higher latency of PMem.
In fact, as we show later, there is a $\approx$\SI{10}{\percent} overhead for accessing data when DRAM acts as a L4 cache instead of normally.

Because it is not possible to leverage the persistency of PMem in memory mode, we focus on \textbf{app direct mode} in the remainder of this paper.
App direct mode, unlike memory mode, leaves the regular memory system untouched.
It optionally allows programs to make use of PMem in the form of memory mapped files.
We describe this process from a developer point of view in the following:

We are using a two-socket system with $24$ physical ($48$ virtual) cores on each node.
The machine is running Fedora with a Linux kernel version 4.15.6.
Each socket has \SI{6}{\textrm{PMem}} DIMMs with \SI{128}{\giga\byte} each and \SI{6}{\textrm{DRAM}} DIMMs with \SI{32}{\giga\byte} each.

To access PMem, the physical PMem DIMMs first have to be grouped into so-called \textit{regions} with \texttt{ipmctl}\footnote{\textbf{\texttt{ipmctl:}} https://github.com/intel/ipmctl}:
\begin{lstlisting}[]
ipmctl create -f -goal -socket 0 MemoryMode=0 \
PersistentMemoryType=AppDirect
\end{lstlisting}
To avoid complicating the following experiments with a discussion on NUMA effects (which are similar to the ones on DRAM) we run all our experiments on \texttt{socket 0}.
Once a region is created, \texttt{ndctl}\footnote{\textbf{\texttt{ndctl:}} https://github.com/pmem/ndctl} is used to create a namespace on top of it:
\begin{lstlisting}[]
ndctl create-namespace --mode fsdax --region 28
\end{lstlisting}
Next, we create a file system on top of this namespace (\texttt{mkfs.ext4}\footnote{\textbf{\texttt{mkfs.ext4:}} https://linux.die.net/man/8/mkfs.ext4}) and mount it (\texttt{mount}\footnote{\textbf{\texttt{mount:}} https://linux.die.net/man/8/mount}) using the \texttt{dax} flag, which enables direct cache-line-grained access to the device by the CPU:
\begin{lstlisting}[]
mkfs.ext4 /dev/pmem28
mount -o dax /dev/pmem28 /mnt/pmem28/
\end{lstlisting}
Programs can now create files on the newly mounted device and map them into their address space using \texttt{mmap}\footnote{\textbf{\texttt{mmap:}} http://man7.org/linux/man-pages/man2/mmap.2.html}:
\begin{lstlisting}[]
fd = open(("/mnt/pmem28/file", O_RDWR, 0);
res = ftruncate(fd, SIZE);
ptr = mmap(nullptr, SIZE, PROT_WRITE, MAP_SHARED, fd, 0);
\end{lstlisting}

The pointer can be used to access the PMem directly, just like regular memory.
\Cref{sec:io_primitives} discusses how to ensure that a value written to PMem is actually persistent.
In the remainder of this section, we discuss the bandwidth and latency of PMem.


\subsection{Bandwidth}
\label{sec:bandwidth}

It is important to know that the PMem hardware works internally on \SI{256}{\textrm{byte}} blocks.
A small write-combining buffer is used to avoid write amplification, because the transfer size between PMem and CPU is, as for DRAM, \SI{64}{\textrm{byte}} (cache lines).

The block-based (\SI{4}{\textrm{cache lines}}) design of PMem leads to some interesting performance characteristics that we show in \Cref{fig:bandwidth_scale_cls}.
The experiment measures the bandwidth for loading/storing from/to independent random locations on PMem and DRAM.
We use all \SI{24}{\textrm{physical cores}} of one socket to maximize the number of parallel accesses.
The figure shows store (PMem: (\textbf{a}), DRAM: (\textbf{b})) and load (PMem: (\textbf{c}), DRAM: (\textbf{d})) benchmarks.
The performance depends significantly on the number of consecutively accessed cache lines on PMem, while there is no significant difference on DRAM.
Peak throughput can only be reached when a multiple of the block size (\SI{4}{\textrm{cache lines}} = \SI{256}{\textrm{byte}}) is used.

As on DRAM, streaming (non-temporal) stores are more efficient on PMem because the modified cache lines do not have to be loaded first---thereby saving memory bandwidth.
However, on PMem the performance of regular stores can be increased to that of streaming stores by issuing a \texttt{clwb} (cache line write back) instruction after each store.
The \texttt{clwb} forces a dirty cache line in the data cache to be written to the underlying memory system (without evicting the cache line).
While this is beneficial on PMem \textbf{(a)}, it does not change the throughput on DRAM \textbf{(b)}.

This effect is studied further in \Cref{fig:bandwidth_scale_threads}, which shows the same experiment, but instead of varying the number of cache lines loaded/stored we vary the number of threads.
It shows that the \texttt{clwb} instruction only becomes necessary once several threads are writing to PMem:
With more threads, cache lines are evicted more randomly from the last level CPU cache, and thus arrive increasingly out of order at the PMem write-combining buffer.
It seems that at a certain point ($\approx4$ threads), the buffer is no longer able to combine the cache lines into a single PMem block write.
Using the \texttt{clwb} instruction, we can force the order in which the cache lines arrive at the PMem write buffer and thus enable it to combine neighboring cache lines into a single block write.

Another effect we observe is that the throughput peaks at around \SI{3}{\textrm{threads}} for streaming (and 12 for stores with \texttt{clwb}).
Using additional threads decreases the throughput slightly.

Lastly, a largely unrelated but somewhat amusing effect of the hardware pre-fetcher is shown in \Cref{fig:bandwidth_scale_cls} \textbf{(c)} and \textbf{(d)}.
Starting at $10$ adjacent cache lines, the pre-fetcher becomes active and fetches additional cache lines.
If these are not needed, as in our experiment, the effective throughput suffers.

In summary, judging from our experimental results, we recommend the following guidelines for bandwidth-critical applications:
\begin{itemize}
	\item Algorithms should no longer be designed to fit data on single cache lines (\SI{64}{\textrm{byte}}) but on PMem blocks (\SI{256}{\textrm{byte}}).
	\item Streaming operations should be utilized when possible, otherwise stores should be followed by \texttt{clwb}.
	\item Over-saturating PMem can lead to reduced performance.
	\item The experiments showed that the PMem read bandwidth is $2.6\times$ lower and the write bandwidth $7.5\times$ lower than DRAM.
	Therefore, performance-critical code should prefer DRAM over PMem (e.g., by buffering writes in a DRAM cache).
\end{itemize}


\subsection{Latency}
\label{sec:latency}

While bandwidth is critical for OLAP-style applications, latency is much more important for OLTP workloads because the access pattern shifts from large scan operations to (sequential I/O) to point lookups, which are essentially random accesses into memory.
The performance of these random accesses is dominated by the latency of the underlying device.

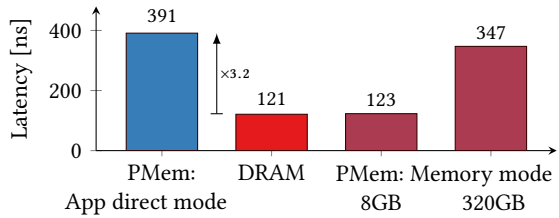
\begin{figure}
\begin{tikzpicture}
\begin{axis}[
axis lines=left,
ymin=0, ymax=480,
xmin=0.4, xmax=4.6,
xtick={1,2,3,4},
height=3.5cm,
width=\linewidth/1.1,
bar width=3em,
xticklabel style={align=center},
xticklabels={PMem:\\ \hspace{-0.5cm} App direct mode, DRAM, \hspace{1.6cm} PMem: Memory mode\\ 8GB,\vphantom{[]}\\{} 320GB},
ylabel={Latency [\SI{}{\nano\second}]}
];

\addplot[ybar, red!20!black, fill=color1] plot coordinates {%
	(1, 391)
};
\node[align=center, text width=3cm] at (1,450) {\small 391};

\addplot[ybar, red!20!black, fill=color2] plot coordinates {%
        (2, 121)
};
\node[align=center, text width=3cm] at (2,172) {\small 121};

\addplot[ybar, red!20!black, fill=colorMix1] plot coordinates {%
        (3, 123)
};
\node[align=center, text width=3cm] at (3,170) {\small 123};

\addplot[ybar, red!20!black, fill=colorMix1] plot coordinates {%
	(4, 347)
};
\node[align=center, text width=3cm] at (4,394) {\small 347};

\draw [|-latex] (1.5,121) -> (1.5,391) node[midway, text width=.5cm,text opacity=1,rounded corners, inner sep=2, xshift=0.3cm] {\tiny $\mathbf{\times 3.2}$} {};

\end{axis}
\end{tikzpicture}

\beforeCaptionSpacing
\caption{\textbf{Read Latency} -- Random access read latency.}
\afterCaptionSpacing
\label{fig:latency}
\end{figure}

To measure the latency for load operations on PMem, we use a single thread and perform loads from random locations.
To study this effect, we prevented out-of-order execution by chaining the loads such that the address for the load in step $i$ depends on the value read in step $i-1$.
The results are shown in \Cref{fig:latency}.

We can observe that DRAM read latency is lower than PMem by a factor of $3.2$.
Note that this does not mean that each access to PMem is that much slower, because many applications can still benefit from the regular on-CPU L3 cache.
When PMem is used in memory mode, it replaces DRAM as main memory and DRAM acts as an L4 cache.
In this configuration, the data size is important:
When using \SI{8}{\giga\byte} (as in the other modes) the performance is similar to that of DRAM, because the DRAM cache captures all accesses.
However, when we increase the data size to \SI{360}{\giga\byte}, the DRAM cache (around \SI{200}{\giga\byte} on the socket we use) is not hit that frequently and the performance degrades.

\begin{figure}
\begin{tikzpicture}
\begin{axis}[
axis lines=left,
ymin=0, ymax=990,
xmin=0.3, xmax=14.7,
height=3.5cm,
width=\linewidth/1.1,
bar width=0.5em,
xtick={1,2,3,4,5,6,7,8,9,10,11,12,13,14},
xticklabel style={xshift=-5},
xticklabels={,,Single,,,,,Sequential,,,,,Random,,},
ylabel={Latency [\SI{}{\nano\second}]},
legend columns=2,
];
\legend{flush, flush\_opt, clwb, streaming}


\addplot[ybar, red!20!black, fill=color1, area legend] plot coordinates {(1, 750)};
\addplot[ybar, red!20!black, fill=color3, area legend] plot coordinates {(2, 808)};
\addplot[ybar, red!20!black, fill=color4, area legend] plot coordinates {(3, 811)};
\addplot[ybar, red!20!black, fill=color2, area legend] plot coordinates {(4, 183)};

\addplot[ybar, red!20!black, fill=color1] plot coordinates {(6, 108)};
\addplot[ybar, red!20!black, fill=color3] plot coordinates {(7, 104)};
\addplot[ybar, red!20!black, fill=color4] plot coordinates {(8, 104)};
\addplot[ybar, red!20!black, fill=color2] plot coordinates {(9, 174)};

\addplot[ybar, red!20!black, fill=color1] plot coordinates {(11, 125)};
\addplot[ybar, red!20!black, fill=color3] plot coordinates {(12, 130)};
\addplot[ybar, red!20!black, fill=color4] plot coordinates {(13, 127)};
\addplot[ybar, red!20!black, fill=color2] plot coordinates {(14, 174)};


\node[right] at (0.33, 810) {\footnotesize 750};
\node[right] at (1.33, 868) {\footnotesize 808};
\node[right] at (2.33, 871) {\footnotesize 811};
\node[right] at (3.33, 243) {\footnotesize 183};

\node[right] at (5.33, 168) {\footnotesize 108};
\node[right] at (6.33, 164) {\footnotesize 104};
\node[right] at (7.33, 164) {\footnotesize 104};
\node[right] at (8.33, 234) {\footnotesize 174};

\node[right] at (10.33, 185) {\footnotesize 125};
\node[right] at (11.33, 190) {\footnotesize 130};
\node[right] at (12.33, 187) {\footnotesize 127};
\node[right] at (13.33, 234) {\footnotesize 174};

\end{axis}
\end{tikzpicture}

\beforeCaptionSpacing
\caption{\textbf{Persistent Write Latency} -- Access latency for writing cache lines persistently.}
\afterCaptionSpacing
\label{fig:latency_write}
\end{figure}
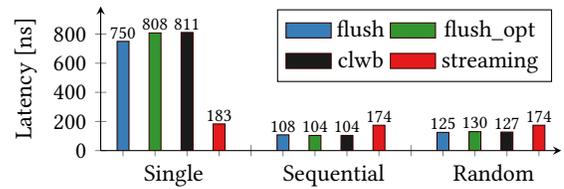

To store data persistently on PMem, the data has to be written, the cache line evicted, and then an \texttt{sfence} has to be used to wait for the data to reach PMem.
This process is described in more detail in \Cref{sec:failure_atomicity}.
To measure the latency for persistent store operations on PMem, we use a single thread that persistently stores data to an array of \SI{10}{\giga\byte} in size.
Each store is aligned to a cache line (\SI{64}{\textrm{byte}}) boundary.
The results are shown in \Cref{fig:latency_write}.

The four bars on the left show the results for continuously writing to the same cache line, in the middle we write cache lines sequentially, and on the right randomly.
In each scenario, we use four different methods for flushing cache lines (from left to right: \texttt{flush}, \texttt{flushopt}, \texttt{clwb}, and streaming stores).

When data is written to the same cache line, streaming stores should be preferred.
This pattern appears in many data structures (e.g., array-like structures with a size field) or algorithms (e.g., a global counter for time-stamping) that have some kind of global variable that is often modified.
Therefore, for efficient usage of PMem, techniques similar to the ones developed to avoid congestion in multi-threaded programming have to be applied to PMem as well.
Among non-streaming instructions, there is no significant difference, because the Cascade Lake CPUs do not fully implement \texttt{clwb}.
Intel has added opcode to allow software to use it, but implement it as \texttt{flush\_opt} for now.
Therefore, streaming operations and \texttt{clwb} should be preferred over \texttt{flush} and \texttt{flush\_opt}.

\section{Storage Primitives for PMem}
\label{sec:io_primitives}

The low write latency of PMem (compared to other storage devices) makes it an ideal candidate for use in database systems, file systems, and other systems software.
However, due to the CPU cache, writes to PMem are only persistent once the corresponding cache line is flushed.
Algorithms have to explicitly order stores and cache line flushes to ensure that a persistent data structure is always in a consistent state (in case of a crash).
We call this property \textit{failure atomicity} and discuss it in \Cref{sec:failure_atomicity}.
Intel's Persistent Memory Development Kit (PMDK)~\cite{PMDK}, an open-source library for Pmem, abstracts from this complexity by providing two failure atomic I/O primitives: log writing (\textit{libpmemlog}) and block/page flushing (\textit{libpmemblk}).
In \Cref{sec:tx_logging} and \Cref{sec:page_flush}, we apply the guidelines developed earlier (\Cref{sec:micro_benchmarks}), apply them to these two problems, and analyze their performance.

\subsection{Failure Atomicity}
\label{sec:failure_atomicity}

As mentioned earlier, when data is written to PMem, stores are not immediately propagated to the PMem device, but instead buffered in the regular on-CPU cache.
While programs cannot prevent the eviction, they can force it using explicit write-back or flush instructions.
This implies that any persistent data structure on PMem always needs to be in a consistent state, otherwise a system crash---interrupting an update operation---could lead to an inconsistent state after a restart.
The following code snippet shows how an element is appended to a pre-allocated buffer:
\\
\begin{minipage}[b]{.35\linewidth}
	\begin{lstlisting}
	struct Buffer {
	  int eles[128];
	  int next;
	};
	
	
	
	
	\end{lstlisting}
\end{minipage}\qquad
\begin{minipage}[b]{.65\linewidth}
	\begin{lstlisting}[xleftmargin=0pt]
	void append(Buffer* buf, int ele) {
	  buf->eles[buf->next] = ele;
	  clwb(&buf->eles[buf->next]);
	  sfence();
	  buf->next++;
	  clwb(&buf->next);
	  sfence();
	}
	\end{lstlisting}
\end{minipage}
The new element is first copied into the next free slot and the corresponding cache line is forced to be written back to PMem.
Instead of using a regular flush operation, \texttt{clwb} (cache line write back) is used, which is an efficient flush operation designed for PMem that flushes the cache line without invalidating it.
Before the buffer's size indicator (\texttt{next}) can be changed, a \texttt{sfence} (store fence) must be issued to prevent re-ordering by the compiler or hardware.
Once \texttt{next} has been written, it is persisted to memory in the same fashion.
Note that persisting the \texttt{next} field is not necessary for the failure atomicity of a single append operation.
However, it is convenient and often required for subsequent code (e.g., another append).
In the following, we will use the term persistency barrier and persist for a combination of a \texttt{clwb} and a subsequent \texttt{sfence}:
\begin{lstlisting}
void persist(void* ptr) { clwb(ptr); sfence(); }
\end{lstlisting}

Generally speaking, a persistency barrier is an expensive operation, as it forces a synchronous write to PMem (or, more precisely, to its internal battery-backed buffers).
Therefore, in addition to the guidelines laid out in \Cref{sec:micro_benchmarks}, it is also important to minimize the number of persistency barriers while still maintaining failure atomicity.
In the following two sections, we show a manually-tuned implementation for logging and page flushing.


\subsection{Page Propagation}
\label{sec:page_flush}

\newcommand{\flushPageChartScale}{2.9}

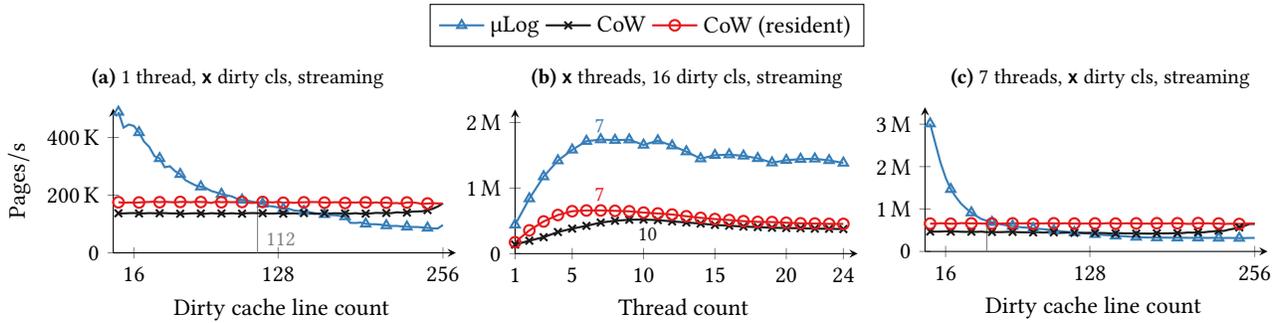
\begin{figure*}
	\centering
	\ref{legend:flush}
	\\
	\vspace{2mm}
	\subcaptionbox{1 thread, \texttt{\textbf{x}} dirty cls, streaming} {
		\hspace{-2mm}
		\begin{tikzpicture}
   \begin{axis}[
      width=\textwidth/\flushPageChartScale,
      height=3.5cm,
      xlabel={Dirty cache line count \vphantom{[]}},
      axis lines=left,
      ymin=0, ymax=500,
      xmin=0, xmax=266,
      legend columns=3,
      legend cell align={left},
      legend to name={legend:flush},
      legend style={at={(0.5,0.8)},anchor=west},
      mark repeat = 4,
      mark phase = 1,
      xtick={16,128,256},
      ylabel={\si{Pages\per\second}},
      yticklabel={%
      	\ifthenelse{\equal{\tick}{0.0}}
      	{0}
      	{\SI[round-mode=places, round-precision=0]{\tick}{K}}
      },
      ]
      \legend{{\textmu}Log, CoW, CoW (resident)}

      \node[right, gray] at (112.0,50) {\small 112};
      \addplot[samples=1, domain=0:6, gray, name path=three, forget plot] coordinates {(112,0)(112,170)};

      \addplot[thick,color=color1,mark=triangle] file {figures/page_flush/st_scale_cl/data/streaming_micro.data};

      \addplot[thick,color=color4,mark=x] file {figures/page_flush/st_scale_cl/data/streaming_shadow.data};

      \addplot[thick,color=color2,mark=o] file {figures/page_flush/st_scale_cl/data/streaming_shadow_resident.data};

   \end{axis}
\end{tikzpicture}
	}
	\hspace{2mm}
	\subcaptionbox{\texttt{\textbf{x}} threads, 16 dirty cls, streaming} {
		\hspace{-8mm}
		\begin{tikzpicture}
   \begin{axis}[
      width=\textwidth/\flushPageChartScale,
      height=3.5cm,
      xlabel={Thread count \vphantom{[]}},
      axis lines=left,
      ymin=0, ymax=2.2,
      xmin=1, xmax=25,
      xtick={1,5,10,15,20,24},
      yticklabel={%
      	\ifthenelse{\equal{\tick}{0.0}}
      	{0}
      	{\SI[round-mode=places, round-precision=0]{\tick}{M}}
      },
      ytick={0,1,2},
      ]

      \addplot[thick,color=color1,mark=triangle] file {figures/page_flush/scale_threads/data/streaming_micro.data};

      \addplot[thick,color=color4,mark=x] file {figures/page_flush/scale_threads/data/streaming_shadow.data};

      \addplot[thick,color=color2,mark=o] file {figures/page_flush/scale_threads/data/streaming_shadow_resident.data};

      \node[right, color1] at (6.0,1.95) {\small 7};
      \node[right, color4] at (9.0,0.3) {\small 10};
      \node[right, color2] at (6.0,0.9) {\small 7};

   \end{axis}
\end{tikzpicture}
	}
	\hspace{4mm}
	\subcaptionbox{7 threads, \texttt{\textbf{x}} dirty cls, streaming} {
		\hspace{-8mm}
		\begin{tikzpicture}
   \begin{axis}[
      width=\textwidth/\flushPageChartScale,
      height=3.5cm,
      xlabel={Dirty cache line count \vphantom{[]}},
      axis lines=left,
      ymin=0, ymax=3.4,
      xmin=0, xmax=266,
      mark repeat = 4,
      mark phase = 1,
      xtick={16,128,256},
      yticklabel={%
      	\ifthenelse{\equal{\tick}{0.0}}
      	{0}
      	{\SI[round-mode=places, round-precision=0]{\tick}{M}}
      },
      ]

      \addplot[samples=1, domain=0:6, gray, name path=three, forget plot] coordinates {(48,0)(48,0.6)};

      \addplot[thick,color=color1,mark=triangle] file {figures/page_flush/mt_scale_cl/data/streaming_micro.data};

      \addplot[thick,color=color4,mark=x] file {figures/page_flush/mt_scale_cl/data/streaming_shadow.data};

      \addplot[thick,color=color2,mark=o] file {figures/page_flush/mt_scale_cl/data/streaming_shadow_resident.data};

   \end{axis}
\end{tikzpicture}
	}
\beforeCaptionSpacing
\caption{\textbf{Failure Atomic Page Flush} -- Flushing \SI{16}{\kilo\byte}~pages (\SI{256}{\textrm{cache lines}} each) in a failure atomic way from DRAM to PMem.}
\afterCaptionSpacing
\label{fig:page_flush}
\end{figure*}

Besides logging, the other essential storage engine component that requires I/O is the buffer manager.
It is responsible for loading (swapping in) pages from SSD/HDD into DRAM whenever a page is accessed by the query engine.
When the buffer pool is full, the buffer manager needs to evict pages in order to serve new requests.
When a dirty page is evicted and has been modified, it needs to be flushed to storage before it can be dropped from the buffer pool, in order to ensure durability.
This process has to be carefully coordinated with the transaction and logging controller, i.e., a page can only be flushed when the undo information of all non-committed modifications is persisted in the log file (otherwise a crash would lead to corrupt data).
In addition, flushing a page needs to be failure atomic: After a crash, the recovery component needs a consistent snapshot of the page.

Flushing pages to persistent storage is an inherently I/O-bound task.
To reduce the latency for pages requests, the buffer manager constantly flushes dirty pages to persistent storage in the background.
This way, it can always serve requests without needing to flush a page first.
In addition, this makes flushing pages (on a background thread) a mostly bandwidth-critical problem (compared to log writing, where latency is most important).

For SSDs/HDDs, this architecture is strictly necessary as pages have to be copied to DRAM before they can be read or written by the CPU.
When PMem is used instead, the buffer pool becomes optional.
However, as recent work~\cite{hymem, arulraj2019multi} has shown, it is still beneficial to use a buffer pool, due to the lower latencies and reduced complexity when working on DRAM compared to PMem.
In addition, this architecture is used in most existing disk-based database systems.
In order to integrate PMem into existing systems, the page flushing algorithm needs to be correct (failure atomicity) and efficient (high bandwidth).
In the following, we describe two algorithms for failure atomic page flushing and then evaluate them.

\subsubsection{Copy-on-Write}


\begin{figure}
   \captionof{lstlisting}{\textbf{Failure Atomicity} -- Pseudo code to flush a DRAM page (\texttt{page\_v}) to a PMem page (\texttt{page\_nv}). CoW: left, {\textmu}Log: right.}
   \label{lst:flush_code}
   \begin{minipage}[b]{.45\linewidth}
      \begin{lstlisting}[language=c, numbers=left, frame=none]
         // 1. Write data
         page_nv.data = page_v.data;
         persist(page_nv);
         // 2. Make PMem page valid
         page_nv.pid = page_v.pid;
         sfence();
         page_nv.pvn = page_v.pvn;
         persist(page_nv.pid
               , page_nv.pvn);



      \end{lstlisting}
   \end{minipage}\qquad
   \begin{minipage}[b]{.45\linewidth}
      \begin{lstlisting}[language=c]
         // 1. Invalidate (*\textcolor{greencomments}{\textbf{\textmu}}*)log
         (*\textmu*)log.pid = INVALID;
         persist(log.pid);
         // 2. Write to (*\textcolor{greencomments}{\textbf{\textmu}}*)log
         (*\textmu*)log <- page_v.dirty_cls
         persist((*\textmu*)log);
         // 3. Set (*\textcolor{greencomments}{\textbf{\textmu}}*)log valid
         (*\textmu*)log.pid = page_v.pid;
         persist((*\textmu*)log.pid);
         // 4. Write to page
         page_nv <- page_v.dirty_cls
         persist(page_nv);
      \end{lstlisting}
   \end{minipage}
\end{figure}

\textbf{CoW} does not overwrite the original PMem page, but instead writes the DRAM page to an unused PMem page~\cite{DBLP:conf/sigmod/ArulrajPD15} (left-hand side of \Cref{lst:flush_code}, line 1-3).
Once the new PMem page is persisted, it is marked as valid (line 5-10) and the old PMem page can be reused.
During recovery, the headers of all PMem pages are inspected to determine the physical location of each logical page.
By adding a page version number (\textit{pvn}) that is increased after each flush, we can identify the latest version of a page.
Using the \textit{pvn}, it becomes unnecessary to invalidate the old PMem page before writing the new one.
This lowers the number of required persistency barriers from three to two and thus yields $\approx10\%$ increased throughput.
We illustrate the \textit{pvn} in the following example:
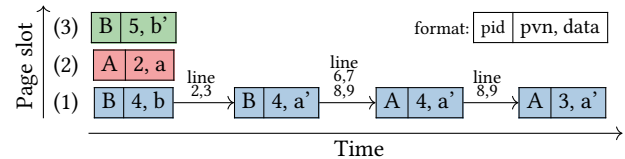
\begin{figure}[H]
\centering
	\vspace{-1mm}

	\begin{tikzpicture}[
		style={},
	]
	
	\setlength{\tabcolsep}{0.3em}

	\node[anchor=south, inner sep=0, outer sep=0, fill=color3!40] at (1.0, 1.0) {
		\begin{tabular}{|l|l|}
		\hline
		B & 5, b' \\ \hline
		\end{tabular}
	};

	\node[anchor=south, inner sep=0, outer sep=0, fill=color2!40] at (1.0, 0.5) {
	\begin{tabular}{|l|l|}
		\hline
		A & 2, a \\ \hline
		\end{tabular}
	};

	\node[anchor=south, inner sep=0, outer sep=0, fill=color1!40] (step1) at (1.0, 0) {
		\begin{tabular}{|l|l|}
		\hline
		B & 4, b \\ \hline
		\end{tabular}
	};
	
	\node[anchor=south, inner sep=0, outer sep=0, fill=color1!40] (step2) at (2.9, 0.0) {
		\begin{tabular}{|l|l|}
		\hline
		B & 4, a' \\ \hline
		\end{tabular}
	};

	\node[anchor=south, inner sep=0, outer sep=0, fill=color1!40] (step3) at (4.8, 0.0) {
		\begin{tabular}{|l|l|}
		\hline
		A & 4, a' \\ \hline
		\end{tabular}
	};

	\node[anchor=south, inner sep=0, outer sep=0, fill=color1!40] (step4) at (6.7, 0.0) {
		\begin{tabular}{|l|l|}
		\hline
		A & 3, a' \\ \hline
		\end{tabular}
	};

	\node[anchor=south] at (5.1, 1.0) {\footnotesize{format:}};
	\node[anchor=south, inner sep=0, outer sep=0] (format) at (6.41, 1.0) {
		\begin{tabular}{|l|l|}
		\hline
		\footnotesize{pid} & \small{pvn, data} \\ \hline
		\end{tabular}
	};

	\draw[->] (step1) to (step2);
	\draw[->] (step2) to (step3);
	\draw[->] (step3) to (step4);

	\draw[->] (-0.2, 0) to (-0.2, 1.5);
	\node at (0.1, 0.2) {(1)};
	\node at (0.1, 0.7) {(2)};
	\node at (0.1, 1.2) {(3)};
	\node[rotate=90] at (-0.42, 0.7) {Page slot};

	\draw[->] (0.4, -0.2) to (7.5, -0.2);
	\node[] at (4, -0.4){Time};

	\node[] at (1.9, 0.55){\footnotesize{line}};
	\node[] at (1.9, 0.35){\footnotesize{2,3}};
	\node[] at (3.8, 0.75){\footnotesize{line}};
	\node[] at (3.8, 0.55){\footnotesize{6,7}};
	\node[] at (3.8, 0.35){\footnotesize{8,9}};
	\node[] at (5.7, 0.55){\footnotesize{line}};
	\node[] at (5.7, 0.35){\footnotesize{8,9}};

	\end{tikzpicture}
	\vspace{-4mm}
\end{figure}

The green page slot \alexNumberNoneRef{3} contains the latest persistent copy of page \texttt{B}.
The red one \alexNumberNoneRef{2} contains the original version of page \texttt{A}.
The different versions of the blue page slot (\alexNumberNoneRef{1}) show each step of flushing a new version of page \texttt{A}.
The line numbers where the transition might occur are written over the arrow.
In each step, the \textit{pvn} can be used to figure out the most recent version of each page.
In database systems, the log sequence number \textit{lsn} could be used instead of the \textit{pvn}, however if the system crashes in line $6$, log entries might be reapplied to a page.

\subsubsection{Micro Log}

The micro-log technique uses a small log file to record changes that are going to be made to the page.
In order to know, which cache lines have been changed, the page is required to track modified areas since its last flush.
During recovery, all valid micro logs are reapplied, independent of the page's state.
This forces us to invalidate the log (right-hand side of \Cref{lst:flush_code}, line 1-3) before changing the content (line 5-7), otherwise the changes would be applied to the previous page in case of a crash.
Only once the changes are written, we set them to valid (line 8-10) and then apply them to the actual page (line 13-15).

\subsubsection{Experiments}

\Cref{fig:page_flush} details the page flush performance.
All techniques are implemented as a micro benchmark using streaming (also known as non-temporal) writes, which have been shown to provide the highest throughput in \Cref{sec:micro_benchmarks}.
When using copy-on-write, we differentiate whether all cache lines are available in DRAM (\tierNoOptLine) or only the dirty ones (\nvmLine).
As a performance metric, we chose the number of pages that can be flushed to PMem per second.
We vary the number of dirty cache lines in \textbf{(a)} for a single thread and in \textbf{(c)} for \num{7} threads.
In \textbf{(b)}, we vary the number of threads to show the scale-out behavior.

The results show that the micro log is efficient when the number of cache lines that have to be flushed is low.
We can observe this effect for a single thread in \textbf{(a)}.
Using the micro log yields performance gains for up to 112 dirty cache lines.
A multi-threaded experiment is shown in \textbf{(c)}.
Here the micro log only offers throughput gains, when fewer than \SI{32}{\textrm{cache lines}} are dirty.
Therefore, a hybrid technique based on a simple cost model should be used to choose the better technique, depending on the number of dirty cache lines (and single/multi threading).

The micro benchmarks in \Cref{sec:micro_benchmarks} suggested that streaming instructions should be preferred over regular stores.
We were able to confirm this finding in the page flushing experiment (not shown in chart).
In addition, as in the bandwidth experiments, we can see a performance degradation when too many threads are used:
For optimal throughput it is important to tailor the number of writer threads to the system.
As \textbf{(b)} shows, the performance degrades after reaching a peak at around 7-11 threads.

\subsection{Logging}
\label{sec:tx_logging}


\begin{figure*}
	\centering
	\ref{ledgend:tx_log}
	\\
	\vspace{2mm}
	\subcaptionbox{Unaligned log entries} {
	\hspace{-8mm}
	\begin{tikzpicture}
	\begin{axis}[
		width=\textwidth/2.1,
		height=3.5cm,
		xlabel={Log entry size [\si{\textrm{byte}}]},
		axis lines=left,
		ymin=0, ymax=10.8,
		xmin=56, xmax=522,
		xtick={56,128,256,384,512},
        mark repeat = 2,
		mark phase = 1,
		legend columns=4,
		legend cell align={left},
		legend to name={ledgend:tx_log},
		legend style={at={(0.5,0.8)},anchor=west},
		ylabel style={align=center},
		ylabel={Throughput\\{} [Log Entries / s]},
		yticklabel={%
			\ifthenelse{\equal{\tick}{0.0}}
			{0}
			{\SI[round-mode=places, round-precision=0]{\tick}{M}}
		},
	]
	\legend{Zero, Header, Classic};
	
	\addplot[thick,color=color1,mark=triangle] file {figures/tx_log/data/unaligned_zero.data};
	\addplot[thick,color=color4,mark=x] file {figures/tx_log/data/unaligned_header.data};
	\addplot[thick,color=color2,mark=o] file {figures/tx_log/data/unaligned_classic.data};
	
	\end{axis}
\end{tikzpicture}
	}
	\label{fig:tx_log_left}
	\hspace{2em}
	\subcaptionbox{Aligned log entries} {
		\hspace{-8mm}
		\begin{tikzpicture}
	\begin{axis}[
		width=\textwidth/2.1,
		height=3.5cm,
		xlabel={Log entry size [\si{\textrm{byte}}]},
		axis lines=left,
		ymin=0, ymax=10.8,
		xmin=56, xmax=522,
		xtick={56,128,256,384,512},
        mark repeat = 2,
		mark phase = 1,
		legend columns=4,
		legend cell align={left},
		legend to name={ledgend:tx_log},
		legend style={at={(0.5,0.8)},anchor=west},
		yticklabel={%
			\ifthenelse{\equal{\tick}{0.0}}
			{0}
			{\SI[round-mode=places, round-precision=0]{\tick}{M}}
		},
	]
	\legend{Zero, Header, HeaderDance, Classic};
	
	\addplot[thick,color=color1,mark=triangle] file {figures/tx_log/data/aligned_zero.data};
	\addplot[thick,color=color4,mark=x] file {figures/tx_log/data/aligned_header.data};
	\addplot[thick,color=color3,mark=square] file {figures/tx_log/data/aligned_dance_header.data};
	\addplot[thick,color=color2,mark=o] file {figures/tx_log/data/aligned_classic.data};

	\end{axis}
\end{tikzpicture}
	}
	\label{fig:tx_log_right}
	\beforeCaptionSpacing
	\caption{\textbf{Transaction Log} -- The throughput for writing log entries of varying size to PMem.}
	\afterCaptionSpacing
	\label{fig:tx_log}
\end{figure*}
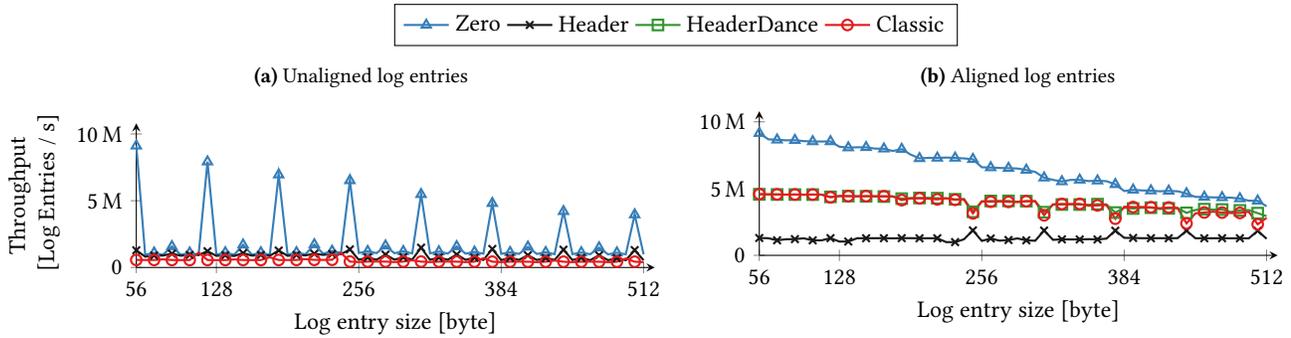

In database systems, write-ahead logging is used to ensure the atomicity and durability of transactions.
This is achieved by recording (logging) the individual changes of a larger transaction in order to be able to undo them in the event of a rollback.
If any of the changes to the data are persisted while the transaction is still active, the log has to be persisted as well.
Before a transaction is completed (thereby guaranteeing to the user, that all changes of the transaction are durable), all log entries of the transaction are written persistently.
Logging allows a database to only persist the delta of the modifications:
For example, consider an insert into a table stored as a B-Tree: Using logging, only the altered data needs to be persisted instead of all modified nodes (pages).
During restart, the recovery component reads the log file, determines the most recent fully persisted log entry, and applies the log to the database.

Logging constitutes a major performance bottleneck in database systems using traditional storage devices (SSD/HDD) because each transaction has to wait until the log entry recording its changes is written.
As a mitigation, reduced consistency guarantees are offered and complex group commit protocols are implemented.
However, using PMem, a low-latency logging protocol can be implemented that largely eliminates this problem.

\subsubsection{Algorithms}

In the following, we first explain and then evaluate three logging techniques: \textit{Classic}, \textit{Header}, and \textit{Zero}:

\textbf{Classic} represents a form of logging commonly used in database systems~\cite{DBLP:phd/dnb/Sauer17}.
The following listing shows the algorithm in pseudo code (left) and the file layout grammar (right).
For clarity, only information relevant to the protocol is depicted.
\\
\begin{minipage}[b]{.55\linewidth}
	\begin{lstlisting}
	log << header << payload
	persist(log);
	log << footer
	persist(log);
	\end{lstlisting}
\end{minipage}\qquad
\begin{minipage}[b]{.45\linewidth}
	\begin{lstlisting}[xleftmargin=0pt]
	LogFile -> Entry*
	Entry -> header (*$\hookleftarrow$*)
	          payload footer
	
	\end{lstlisting}
\end{minipage}

A log entry is flushed in two steps: First, the header and payload is appended to the log and persisted; second, the footer, which contains a copy of the log sequence number (lsn; an id given to each log entry).
The lsn in the footer can be used during recovery to determine whether a log entry was completely written and therefore should be considered as valid and applied to the database.
Note that it takes two persistency barriers.
Without the first barrier, parts of the payload could be missing even if the footer is present in PMem, due to the flushes being reordered.

\textbf{Header} uses the same technique as \textit{libpmemlog} in the PMDK~\cite{PMDK}.
It is similar to appending elements to an array:
\\
\begin{minipage}[b]{.55\linewidth}
	\begin{lstlisting}
	log << header << payload
	persist(log);
	log.size += entry_size
	persist(log.size);
	\end{lstlisting}
\end{minipage}\qquad
\begin{minipage}[b]{.45\linewidth}
	\begin{lstlisting}[xleftmargin=0pt]
    LogFile -> size, Entry*
    Entry -> header payload
	
	
	\end{lstlisting}
\end{minipage}
The log entry is also written in two steps: First, the header and payload are appended to the tail of the log and persisted.
Next, the new size of the log is set in the header of the log file and persisted.
This eliminates the need to scan the log file for the last valid entry during recovery because the valid size is directly stored in the header.

\textbf{Zero} is a novel technique we propose for PMem that requires only one persistency barrier:
\\
\begin{minipage}[b]{.55\linewidth}
	\begin{lstlisting}
	cnt = pop_count(header, payload)
	log << header << cnt << payload
	persist(log);
	\end{lstlisting}
\end{minipage}\qquad
\begin{minipage}[b]{.45\linewidth}
	\begin{lstlisting}[xleftmargin=0pt]
	LogFile -> Entry*
	Entry -> header (*$\hookleftarrow$*)
	          pop_cnt payload
	\end{lstlisting}
\end{minipage}
\\
Before logging starts, each log file is initialized to zero.
This is commonly done anyway by database systems (e.g., PostgreSQL) to enforce that the file system actually allocates pages to the file.
When a log entry is written, the number of set bits are counted (using the \texttt{popcnt} instruction).
Next the header, data, and bit count (\texttt{cnt}) is written to the log and persisted together.
Using the bit count, it is always possible to determine the validity of a log entry: Either the cache line containing the bit count was not flushed or it was.
In the former case, the field contains the number zero (because the file was zeroed) and the entry is invalid.
In the latter case, the bit count field can be used to determine whether all other cache lines belonging to the log entry have been flushed as well.

\subsubsection{Experiments}

In \Cref{sec:latency}, we showed that there is a large performance penalty when the same cache line is persisted twice in a row.
This effect is very relevant for latency-critical systems, as shown in \Cref{fig:tx_log}.
We use a micro-benchmark that measures the throughput of flushing log entries of varying sizes.
The left chart shows a naive implementation, while the right one uses padding on each log entry to align  entries to cache line boundaries and thus avoid subsequent writes to the same cache line.
While padding wastes some memory\footnote{Up to \SI{1}{\textrm{cache line}} for \textit{Zero} and \textit{Header}; up to \SI{2}{\textrm{cache lines}} for \textit{Classic}}, the throughput greatly increases ($\approx8\times$).

However, even with padding, the \textit{Classic} approach still outperforms the \textit{Header} one, because of the slowdown due to the writes to the same cache line in the header when the size is updated.
This problem can be solved by using a \textit{dancing} size field:
We use several size fields on different cache lines in the header and only write one (round-robin) for each log entry.
By using \num{64} of these dancing size fields, the throughput of \textit{Header} can be increased to that of \textit{Classic}.
However, both of these techniques still require persistency barriers and therefore cannot compete with \textit{Zero} logging ($\approx2\times$ faster).

The log implementation (\textit{libpmemlog}) of the PMDK~\cite{PMDK} uses the same approach (and therefore yields the same throughput, when locking is disabled) as our naive \textit{Header} implementation without alignment and dancing.
It has the advantage that the log file is dense and can be presented to the user as one continuous memory segment.
However, this leaves the user with the task of reconstructing log entry boundaries manually.
By moving this functionality into the library, a better logging strategy can be implemented and the usability increased.

For validation, we have integrated all techniques into our storage engine prototype HyMem~\cite{hymem}.
Running a write-heavy (100\%) YCSB benchmark~\cite{cooper2010benchmarking} on a single thread with a DRAM-resident table, \textit{Zero} logging, \textit{Header}, and \textit{Classic} achieves a throughput of \num{2}\,M, 1.7\,M, and 1.5\,M transactions per second, respectively.

\section{Related Work}
\label{sec:related}

With PMem only being released recently, this is one of the two~\cite{DBLP:journals/corr/abs-1903-05714} initial studies that have been performed on the actual hardware.
While our work proposes low-level optimizations, Swanson et al. evaluate PMem with various storage engines as well as file systems.
Until now, software or hardware-based simulations, or emulations based on speculative performance characteristics, have been used to evaluate possible system architectures~\cite{DBLP:conf/sigmod/ArulrajP17, DBLP:journals/pvldb/OukidBLLWG17, DBLP:conf/sigmod/OukidL17, DBLP:journals/dbsk/OukidL18}.
The number of persistent index structures~\cite{DBLP:conf/fast/VenkataramanTRC11, DBLP:conf/fast/YangWCWYH15, DBLP:journals/pvldb/ChenJ15, DBLP:conf/fast/LeeLSNN17, DBLP:journals/pvldb/ArulrajLML18, DBLP:conf/icde/Gotze0S18} is large, and has been summarized by G\"otze et.~al~\cite{DBLP:journals/dbsk/GotzeRLLO18}.
Similar techniques have been used to build storage engines directly on PMem~\cite{DBLP:conf/damon/OukidBLBW14, DBLP:conf/sigmod/ArulrajPD15}.
These approaches use in-place updates on PMem, which suffers from the lower-than-DRAM performance.
Therefore, a number of indexes~\cite{DBLP:conf/sigmod/OukidLNWL16, DBLP:conf/usenix/XiaJXS17} as well as storage engines~\cite{DBLP:conf/sigmod/DoZPDNH11, DBLP:journals/pvldb/CanimMBRL10, DBLP:journals/vldb/KangLM16, DBLP:journals/pvldb/LiuS13, DBLP:journals/pvldb/LuoLMCZ12, DBLP:journals/pvldb/AndreiLRSTBMSSV17, DBLP:conf/hpca/KarnagelDRLLSL14} integrate PMem as a separate storage layer or an extension to the recovery component~\cite{DBLP:conf/cidr/OukidLKWB15, DBLP:conf/vldb/OukidNBLBW17}.
Furthermore, buffer-managed architectures~\cite{hymem, arulraj2019multi, DBLP:conf/sigmod/Kimura15} have been proposed to use PMem more adaptively.
Recovery has always been an essential (and performance-critical) component of database systems~\cite{DBLP:phd/dnb/Sauer17}.
Several designs have been proposed for database-specific logging~\cite{DBLP:journals/pvldb/ArulrajPP16, DBLP:conf/icde/FangHHMW11, DBLP:journals/pvldb/HuangSQ14, DBLP:journals/pvldb/WangJ14, DBLP:journals/pvldb/PelleyWGB13} and file systems~\cite{dulloor2014system}.

\section{Conclusion}
\label{sec:conclusion}

In our evaluation, we found several guidelines for using PMem efficiently (cf. \Cref{sec:bandwidth} and \ref{sec:latency}):
(1) Instead of optimizing for cache lines (\SI{64}{\textrm{byte}}) as on DRAM, we have to optimize for PMem blocks (\SI{256}{\textrm{byte}}).
(2) As in multi-threaded programming, writes to the same cache line in close temporal proximity should be avoided.
(3) Forcing the data out of the on-CPU cache (\texttt{clwb} or streaming) is essential for a high write bandwidth.
Furthermore, we evaluated algorithms for logging and page propagation:
(1) Our logging experiments have shown that latency-critical code should minimize the number of persistency barriers and avoid subsequent writes to the same cache line.
(2) Our \textit{zero} logging algorithm reduces the required persistency barriers from two to one, thus doubling the throughput.
(3) For flushing database pages, a small log ({\textmu}Log) can be used to flush only dirty cache lines.
The I/O primitives introduced use an interface similar to the one in PMDK~\cite{PMDK}, making them widely applicable.

\balance
\bibliographystyle{abbrv}
\bibliography{damon}

\end{document}